# Alleviating State-space Explosion in Component-based Systems with Distributed, Parallel Reachability Analysis Algorithm

Vasumathi Narayanan

**Abstract**

In this work, we alleviate the well-known *State-Space Explosion (SSE)* problem in *Component Based Systems (CBS)*. We consider CBS that can be specified as a system of *n Communicating Finite State Machines (CFSMs)* interacting by *rendezvous/handshake* method. In order to avoid the SSE incurred by the traditional product machine composition of the given input CFSMs based on interleaving semantics, we construct a *sum machine* composition based on state-oriented partial-order semantics. The sum machine consists of a set of *n* unfolded CFSMs. By storing statically, just a small subset of global state vectors at *synchronization points*, called the *synchronous environment vectors* and generating the rest of the global-state vectors dynamically on need basis depending on the reachability to be verified, the sum machine alleviates the SSE of the product machine. We demonstrate the implementation of checking the reachability of global state vector from the checking of local reachabilities of the components of the given state vector, through a parallel, distributed algorithm. Parallel and distributed algorithms to generate the sum machine and verifying the reachability in it both without exponential complexity are the contributions of this work.

## 1. Introduction

CBS design is a method of constructing systems with multiple benefits, particularly decreasing the complexity of system design[1]. Model-checking CBS is one of the reliable methods that automatically and systematically analyses the functional correctness of a given system. Nevertheless, model checking is limited by a critical problem so-called Sate Space Explosion (SSE). To benefit from model checking, appropriate methods to reduce SSE, is required. In the last two decades, a number of methods[7], [8], [9] to mitigate the state space explosion have been proposed but they all incur in the worst case, exponential complexity (in the number of components) of the reachability analysis or some heuristics based.

The computational model of the system is conceptual to represent and describe a system. In model checking, a system model originally is represented by Kripke structure, but it can use other graph like representation such as state chart, and Petri net [2]. Using graph like representation by individual states is one of the main representation paradigms in model checking so-called *explicit-state model checking*. Another representation paradigm is *implicit model checking*. In implicit model checking states are not individually represented, but a quantified propositional logic formula is used to represent the graph. In this paper, we focus on only explicit-state model-checking in CBS.

SSE problem is a bottleneck in model checking. The amount of a system's state space (even a finite system) strongly depends on its components and prone to increase in size exponentially. Consequently, it quickly exceeds the memory capacity of the computer and restricts the size that a model checker can check. In the review paper [1] ,the state-space reduction is classified into many categories. Ours is the combination of *scaling down the state-space* and *divide-and-conquer* approach. The general conclusion for this research is that despite proposing many methods for solving the bottleneck of model checking, the SSE still remains an obstacle in worst case and have not been solved completely yet[1]. We claim in our work in all cases the bottleneck due to SSE is completely alleviated which we rigorously show in our computational model.

In the sequel, section 2 explains the computational model of the CBS, we consider. The computational model and consequently many definitions are the same as the ones we considered in our previous work[3],[4],[5]. Section 3 shows the relationship between sum machine and product machine mathematically and thus more rigorously than any of our previous work cited above. Section 4 explains the reachability analysis of an arbitrary global state-vector using sum machine and discusses the complexity of the distributed, parallel algorithm we propose. Section 5 concludes the paper with some pointers to future work. The Appendix lists the pseudo code of the sum machine generation which is again parallel and distributed refinement of our previous work [3], [4].

## 2. The Computational Model of our CBS

Our model is based on *partial-order semantics* as opposed to *interleaving semantics* which is the cause for exponential SSE in the model. We unfold each of the finite CFSM graphs into infinite computation trees (which can be finitely truncated for reachability analysis beyond some *cut-off states* to be defined in the sequel). Each local state of every unfolded CFSM tree stores its *synchronous environment vector*, whose local component is the given state itself and the (*n-1*) non-local components are the most recent synchronization points from the (*n-1*) non-local unfolded trees. The state node from which a transition occurs is called the *input state* and synonymously *source* state or *predecessor* state. The state to which the transition occurs is called the *output state* and synonymously *destination* state or *successor* state. The synchronous transitions occur pairwise and the corresponding two *synchronous output* states are also known as *synchronization points.*

### 2.1 The CFSMs Specification

The CFSM specification is based on Hoare's CSP model [6]. We assume a set of *n* communicating and non-terminating FSMs. Each CFSM is defined as a 6-tuple:

***Definition 1*** A CFSM $F_i = (s_{0fi}, S_{fi}, A_{fi}, R_{tfi}, Rsync_{fi}, Rsync_{0fi})$ $\forall i \in \{1..n\}$ *where,*

- $S_{fi}$ is the *finite* set of states of CFSM $F_i$, $s_{0fi} \in S_{fi}$ being the *initial state*.
- $A_{fi}$ is the *finite set* of *asynchronous* and *synchronous actions of* $F_i$.
- If $a_{fi} \in A_{fi}$ is a *synchronous action*, the list of indices $[j_1, j_2, ... j_k]$, $k \leq n$ of the partner CFSMs are also specified in the square brackets along with $a_{fi}$ .
- $R_{tfi}$ is a *ternary transition relation* such that: $R_{tfi} \subseteq S_{fi} \times A_{fi} \times S_{fi}$ . In a so-called i-transition $(s_{fi}, a_{fi}, s'_{fi}) \in R_{tfi}$ , $s_{fi}$ is called the *input state* and $s'_{fi}$ the *output* state. An i-transition $(s_{fi}, a_{fi}, s'_{fi}) \in R_{tfi}$ is called *synchronous* if $a_{fi}[j_1, j_2, ..., j_k]$, is a *synchronous action* such that : $\exists$ a set of *j-transitions* $(s_{fj}, a_{fj}, s'_{fj}) \in R_{tfj}$ , $\forall j \in \{j_1, j_2, ... j_k\}$ $j \neq i$ where $a_{fi} = a_{fj}$ and $(s'_{fi}, s'_{fj}) \in Rsync_{fi}$
- $Rsync_{fi} \subseteq S_{fi} \times S_{fj}$ , $i \neq j$, $j \in \{1..n\}$, is a binary relation which relates the *output states* of *synchronous transitions*.
- $Rsync_{0fi} \subseteq Rsync_{fi}$ relates the set of pairs of initial states: $Rsync_{0fi} = \{(s_{0fi}, s_{0fj}), \forall j \in \{1..n\}, i \neq j\}$. All the initial states are assumed to be in pairwise synchrony with each other to begin with.

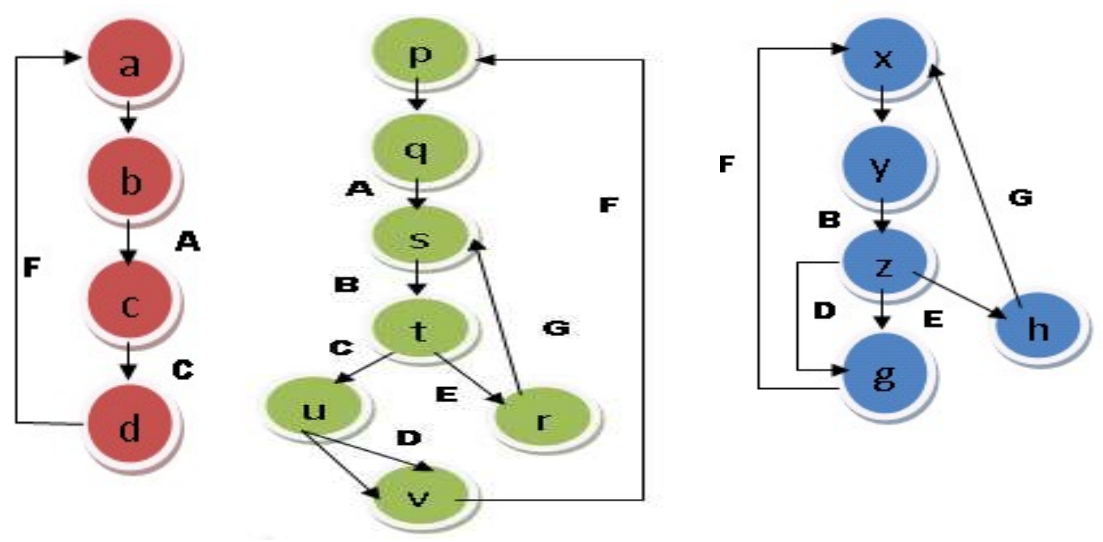

Fig 1. Specification of 3 CFSM Graphs.

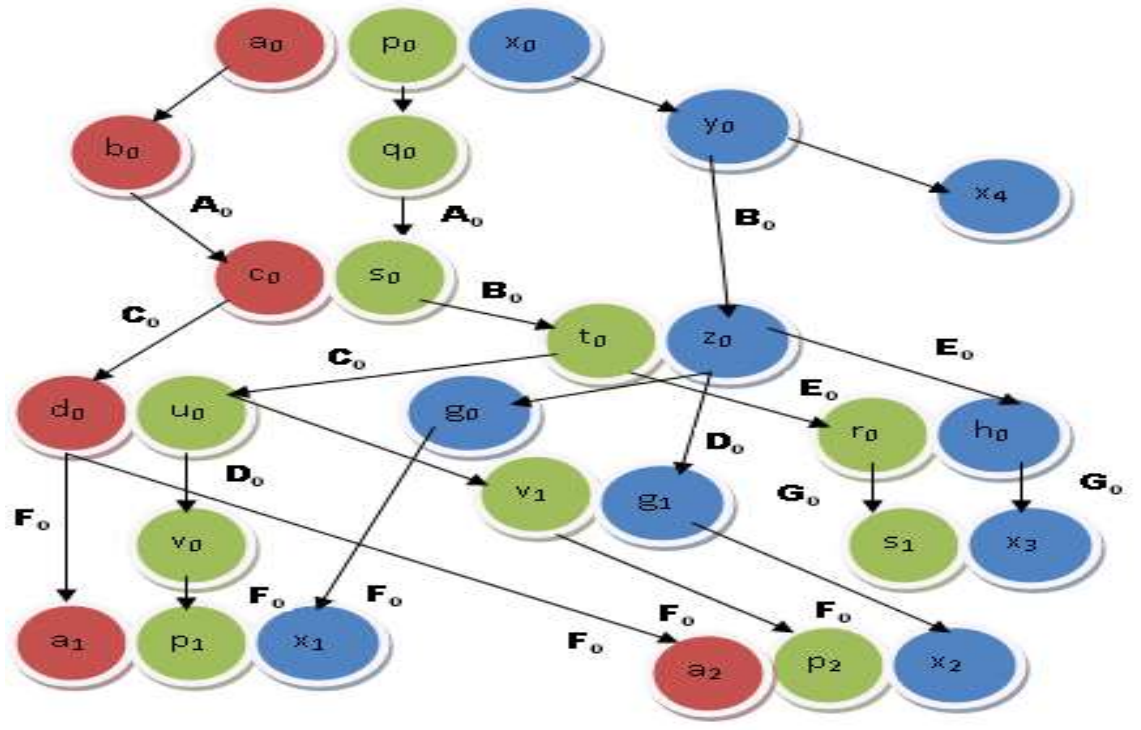

Fig 2. The unfolded set of 3 CFSM Trees.

### 2.2 The Simulation of Non-terminating CFSMs into Finitely Terminating , unfolded CFSMs

The given set of CFSMs represented as *cyclic, rooted, directed graphs* is *simulated in their respective global environments* into a corresponding set of unfoldings, each represented by a *directed, rooted tree* structure.

***Definition.2*** *An unfolded CFSM is a 10-tuple* ($S_i$, $E_i$, $\delta_i$, , $E_{ij}$ , $\delta_{ij}$, , $env_i$, $R_i$, $Rsync_i$, $Rsync_{0i}$ ,$s_{0i}$) $\forall i \in \{1..n\}$ where,

Countably *infinite* sets of states $S_i$ and events $E_i$ of unfolded CFSM $M_i$ are generated as *instances* of corresponding *finite* sets $S_{fi}$ and $A_{fi}$ respectively of CFSM $F_i$, $\forall i \in \{1..n\}$.

$S_i \subseteq S_{fi} \times \mathbb{N}$, $E_i \subseteq A_{fi} \times \mathbb{N}$ *where,*

$\mathbb{N}$ is the set of *natural numbers* with $s_{0i} = (s_{f0i}, 0)$, $\forall i \in \{1..n\}$. Thus,

- $S_i$ is the set of states of unfolding $M_i$,
- $E_i$ is the set of asynchronous events,
- $\delta i : S_i \times E_i \rightarrow S_i$, is the *asynchronous transition function* such that $\delta i(s_i, e_i) = s'_i$ implies that ($s_i$ $R_i$ $s'_i$) where $s_i$ is the *asynchronous input state* and $s'_i$ is the *asynchronous output state*.
- $E_{ij}$ is the set of synchronous events such that for every $e_{ij} \in E_{ij}$, $\exists e_{ji} \in E_{ji}$ such that $e_{ij} = e_{ji}$.
- $\delta_{ij}, : S_i \times S_j \times E_{ij} \rightarrow S_i \times S_j$, is the *synchronous transiton function* such that $\delta_{ij}(s_i, s_j, e_{ij}) = (s'_i, s'_j)$ implies that ($s'_i$ $Rsync_i$ $s'_j$) , ($s_i$ $R_i$ $s'_i$ ) and ($s_j$ $R_j$ $s'_j$ ) where $s_i$, $s_j$ are the *synchronous input states* and $s'_i$, $s'_j$ are the *synchronous output states.*
- $env_i : S_i \rightarrow \times_{k=1..n} S_k$, is the *environment function*, to be explained in a sequel subsection.
- $R_i \subseteq S_i \times S_i$ , $i \in \{1..n\}$ *is the binary relation which relates the input and output states.*
- $Rsync_i \subseteq S_i \times S_j$ , $i \neq j$, $j \in \{1..n\}$, is a binary relation which relates the *output states* of *synchronous transitions*.
- $Rsync_{0fi} \subseteq Rsync_{fi}$ relates the set of pairs of initial states: $Rsync_{0fi} = \{(s_{0fi}, s_{0fj}), \forall j \in \{1..n\}, i \neq j \}$. All the initial states are assumed to be in pairwise synchrony with each other to begin with.
- The initial states are all synchronous output states such that ($s_{i0}$ $Rsync_{0i}$ $s_{j0}$), i={1..n}, i≠j.

### 2.3 Well-founded, Partially-Ordered Causality order among unfolded CFSM states and their Synchronous environment vectors

We unwind the finite CFSM graphs in their mutual global environments into infinite unfolded CFSM *trees* by simulating each of the former in their respective non-local environments.

***Definition 3*** *The global, temporal causality order* is composed using the *binary relations* $Rsync_i$ and $R_i$, where $i \in \{1..n\}$ as follows:

$\leq ::= \quad \cup_{i=\{1..n\}}(R_i \cup Rsync_i)^*$

The binary relation $\leq$ represents the *partially ordered, well-founded causality* relation among the *states* of unfolded CFSMs ordering their *points of entry in time*.

The $Rsync_i$ relations capture the *simultaneity/equality in time* of the *synchronous output states* they relate.

We assume a given specification of *n* CFSMs that are non-terminating. The CFSMs interact by *synchronous message-passing/rendezvous* through pairwise lossless channels, based on the seminal work of CSPs (Communicating Sequential Processes) [10]. Fig 1 shows an example specification of a set of three CFSMs. In Fig 1, *a* is a synchronous action between states *b* to *c* of CFSM $F_1$ and states *q* to *s* of CFSM $F_2$.

## 3. Sequence, Conflict and Concurrency among unfolded CFSM States

***Definition 4*** The three fundamental binary relations *viz. sequence (seq), conflict (conf)* and *concurrency(co) are* defined using $R_i$, $i \in \{1..n\}$ and $\leq$ relations, the latter propagating the local sequence and conflict relations globally across all unfolded CFSMs.

- **Sequence:** $\forall s_i \in S_i, s_j \in S_j$, $(s_i \ seq \ s_j)$ *iff:*

  $\exists \ s'_i \in S_i : (s_i R_i s'_i) \wedge (s'_i \leq s'_j), i, j \in \{1..n\}$, *possibly,* $i=j$

- **Local choice/Conflict:** $\forall s_i, s'_i \in S_i$, $(s_i \ conf_i \ s'_i)$ *iff:* $(s_i, s'_i) \notin R_i^* \wedge (s'_i, s_i) \notin R_i^*, i \in \{1..n\}$

- **Global Choice/Conflict:** $\forall s_i \in S_i, s_j \in S_j, (s_i \ conf \ s_j)$ *iff:* $\exists \ s'_k, s''_k \in S_k : (s'_k \ conf_k \ s''_k) \wedge (s'_k \leq s_i) \wedge (s''_k \leq s_j), i,j,k \in \{1..n\}, i \neq j$, *possibly* $k = i$ *or* $k = j$.

- **Concurrency:** $(s_i \ co \ s_j)$ *iff:* $(s_i, s_j) \notin seq \wedge (s_j, s_i) \notin seq \wedge (s_i, s_j) \notin conf, \ i \neq j$.

- Thus (***seq*** **∪** ***conf*** **∪** ***co***) is a *total* relation.

The above deduction of total relation is a very important result in *state-oriented*, partial-order semantics as opposed to *event-based* systems like Petrinets. In Petrinets the paratial-ordered causality among events captures *happened-before* relation , the complement of which is concurrency relation. Since the *causality* and *concurrency* are complementary, their union is a total relation. This means, *choice/non-determinism* in the specification cannot be modeled on par with sequence and concurrency, sequence being the same as causality.

### 3.1. Properties satisfied by the interaction of Causality, Sequence, Conflict and Concurrency

The causality relation, $\leq$ , defined above is a *partial-order (reflexive, transitive and antisymmetric)*.The sequence relation, *seq,* defined above is *reflexive* (due to *Kleene closure* of *causality*), *transitive* and *asymmetric* which follows from its definition. Conflict relation *conf* and concurrency relation *co* are *irreflexive*, and *symmetric*.

From its definition, it is clear that *sequence* relation is a subset of *causality*, $seq \subseteq \leq$. The concurrency relation, *co* is either unrelated or related by the causality relation, $\leq$. Thus the set ($co \cap \leq$) is a non-empty subset. The members of this subset represent *strong concurrency*, with *necessary co-existence*. On the other hand, those members of co which are unrelated by causality are *weakly concurrent*, which is *possible co-existence*. It is also true that sequence relation *seq* has a non-null intersection with *causality*. That is, ($seq \cap \leq$) is a non empty set.

### 3.2 The Environment state vectors and cut-off vectors

The structure of an unfolded CFSM state along with its synchronous environment vector plays an important role in model-checking. We exploit the fact that ($co \cap \leq$) is a non-empty subset.

***Definition 5*** *The environment* vector $env_i(s_i)$ of an unfolded state $s_i$ is the vector of size $n$, that should n*ecessarily/minimally* be reached according to *causality order* $\leq$ in order to guarantee the entry of state $s_i$ in question. The i[th] component of the environment vector of a state $s_i$ is itself. That is, $env_{ii}(s_i) = s_i$, $\forall i \in \{1..n\}$. The non-local components of the environment vector of a state are all synchronous output states, with $env_i(s_{0i}) = (s_{01}, s_{02}, \ldots s_{0n})$, $\forall i \in \{1..n\}$.
The implication of the above definition is that unfolding of one CFSM calls for the unfolding of the other non-local CFSMs in order to satisfy the possible causal chain of synchronization requirements.

Every state of the unfolded CFSM stores its *synchronous environment state vector*, or simply, the *environment vector*, consisting of its local state, and *(n-1)* non-local components that are synchronous output states. Hence the name synchronous environment state vector. Initial environment vector of all the *n* unfolded component CFSMs is the initial state vector of all the *n* unfoldings. After every *synchronous* transition, the environment vector is updated, with the local component which is a synchronous output state and (*n-1*) non-local components also synchronous output states, that are either most recent *ancestors* or *partners* in the causality order $\leq$. After an *asynchronous* local transition, the *successor* state simply inherits the environment vector of its *predecessor* and more importantly represented independently in its component unfolding as opposed to being interleaved with other independent, non-local asynchronous transitions. Thus we achieve *true concurrency* instead of mimicking it by *non-deterministic interleavings.*

A typical state $a_1$ of $M_1$, the unfolded $CFSM_1$, will have its environment vector as follows:

$$a_1 \underset{co}{\geq} a_2 \underset{co}{\geq} a_3 \underset{co}{\geq} a_4$$

The above example shows that the unfolded $CFSM_1$ state $a_1$ has its environment vector ($a_2$, $a_3$, $a_4$) with number of CFSMs $n$=4, which are mutually related both by causality order $\geq$ and concurrency relation, *co*. $a_2$, $a_3$, $a_4$ are synchronous output states that are members of unfolded CFSMs M2, M3 and M4 respectively. The state $a_4$ must precede $a_3$ which must precede $a_2$ which in turn must precede a1. Because of this chain like causal precedences of $a_1$, the unfolded CFSMs are also known as CMPMs, (*Communicating Minimal Prefix Machines*). Henceforth, the terms unfolded CFSM and CMPM will be used synonymously.

The causality order $\leq$ is the *partial-order*, as opposed to the *total-order* of interleaving semantics. It is a binary relation relating or unrelating all the states of the *n* unfolded computation *trees* of the corresponding CFSM *graphs*. It is derived from the local transition relations $R_i$, $i=1..n$ which accounts for the local *causal precedence* and global synchronous relation *Rsync*, which accounts for the *causal simultaneity* or *strong concurrency*. Uniting the two together, we derive the causality order. It is interesting to note that two states particularly from two different unfolded trees, can be related by causal order and still be concurrent. That is, the causal order need not necessarily mean sequential order of the two states. The $\leq$ relation captures the '*entered before*' order among the two states, while the sequential order *seq* captures the stronger relation than $\leq$, in the sense that if two states $s_i$ and $s_j$ are related by *seq*, $s_i$ has to exit before the entry of $s_j$.

The environment state vectors serve two significant purposes: 1) They form the *statically* saved set of all possible synchronous state vectors that are necessary and sufficient, from which all other unsaved, asynchronous state vectors can be generated *dynamically* on need basis depending on the reachability query of a state vector to be checked. It is precisely the enumeration of asynchronous state vectors which give rise to multiple interleavings, that result in SSE, which we alleviate in our model by storing only the synchronous environment vectors statically and the rest of the state vectors dynamically as demanded by the reachability query. 2) The other purpose is checking the *concurrency* represented by the *co* relation of two different states. The set of environment vectors are necessary and sufficient to generate all other global-state vectors whose reachability needs to be verified, which can be proved by induction trivially by virtue of the generation of unfolded CFSM trees by simulating the given set of CFSM graphs.

### 3.2.1 Cut-off vectors of CMPMs

We define *a function fvec:* $\times_{i=1..n} S_i \rightarrow \times_{i=1..n} S_{fi}$ that maps the unfolded state vectors into corresponding CFSM-state vectors with the instance numbers dropped/omitted.

The *fvec* function is used to map the environment vectors into corresponding CFSM-state vectors, to identify the *cut-off states* of the CMPMs as well.

For example in Fig 2, the state $d_0$ of CMPM $M_1$ (corresponding to state d of CFSM $F_1$) is reachable only if its non-local components $u_0$ and $z_0$ are reachable in addition to its local predecessor $c_0$. Recursively extending this idea, the state $d_0$ of $M_1$ can be reached only if the non-local paths of states $p_0$ to $u_0$ of $M_2$ and $x_0$ to $z_0$ of $M_3$ are reachable in addition to the local path $a_0$ to $c_0$ of itself ($M_1$).

If a given state of a CFSM $F_i$ cannot be reachable say due to *communication deadlock*, it is reflected in its corresponding unfolding CMPM $M_i$. Since we assume finite state systems, eventually as the unfolding proceeds, *fvec(env($s_i$))* repeats. For example, the vector *fvec*($a_0p_0x_0$) = *fvec*($a_1p_3x_2$) = *(apx)*. From then on, the behavior of the CMPM repeats and so we have reached a *cut-off* state-vector, forming a leaf in the CMPM tree. Leaves corresponding to *non-cutoff* state vectors are *dead states* representing a *communication deadlock*.

### 3.2.2 Concurrent path sets of Sum machine

**Definition 5** A *concurrent path set* consists of a set of *n* paths one from each unfolding, such that every pair of states arbitrarily taken from the path set are related either by *seq* relation or *co* relation and there is no *conflict* among them:

$\forall s_i, s'_i, s_j \in S$ it is the case that, $(s_i \; seq \; s'_i) \vee (s_i \; co \; s_j)$, $i \neq j$, $\forall i,j \in \{1..n\}$.

A set of *n finite concurrent paths* $\Pi = (\Pi_1, \Pi_2 ... \Pi_n)$ is given by the set of paths formed by the sequence of state transitions such that $\Pi_i = (s_{0i} R_i s_i R_i ... s'_i)$, $i=\{1..n\}$ and $\forall s_i \in \Pi_i$, $s_j \in \Pi_j$, $\neg(s_i \; conf \; s_j)$ $\forall i,j \in \{1..n\}, i \neq j$.

Recall $(seq \cup conf \cup co)$ is a *total* relation by definition. The set of *n* finite concurrent paths have their *initial configuration* which is given by $(s_{01}, s_{02}, ...s_{0n})$ and *final configuration* given by $(s'_1, s'_2, ...s'_n)$, let us say. A final configuration of a given set of *n* concurrent paths is in general a dynamic configuration. The set of all configurations of a sum machine correspond to the set of all reachable state vectors of the unfolded CFSMs. The set of all *env-vectors* correspond to the set of all *static configurations* of unfolded CFSMs, which is a subset of the set of all reachable configurations.

### 3.2.3 The Sum machine properties of importance in Model-checking

Here we deduce certain properties satisfied by the states of the unfoldings from the definition of various relations above:

**Property 1** Concurrency checking:

$a_i \, co \, b_j$ iff: $env_{ji}(b_j) \; seq \; a_i \wedge env_{ij}(a_i) \; seq \; b_j$.

In other words, $a_i$ is reachable from the i[th] component of environment vector of $b_j$ and similarly $b_j$ can be reached from the j[th] component of the environment vector of $a_i$.

**Proof:** The proof follows from the structure of the states and their respective environment vectors.

The proof has to be in two parts: (i) $(a_i \, co \, b_j) \Rightarrow env_{ji}(b_j) \; seq \; a_i \wedge env_{ij}(a_i) \; seq \; b_j$

(ii) $env_{ji}(b_j) \; seq \; a_i \wedge env_{ij}(a_i) \; seq \; b_j \Rightarrow (a_i \, co \, b_j)$

Consider the following example which consists of proving :

(i) $(a_1 \, co \, b_2) \Rightarrow (b_1 \text{ seq } a_1) \wedge (a_2 \text{ seq } b_2)$

(ii) $(b_1 \; seq \; a_1) \wedge (a_2 \; seq \; b_2) \Rightarrow (a_1 \, co \, b_2)$

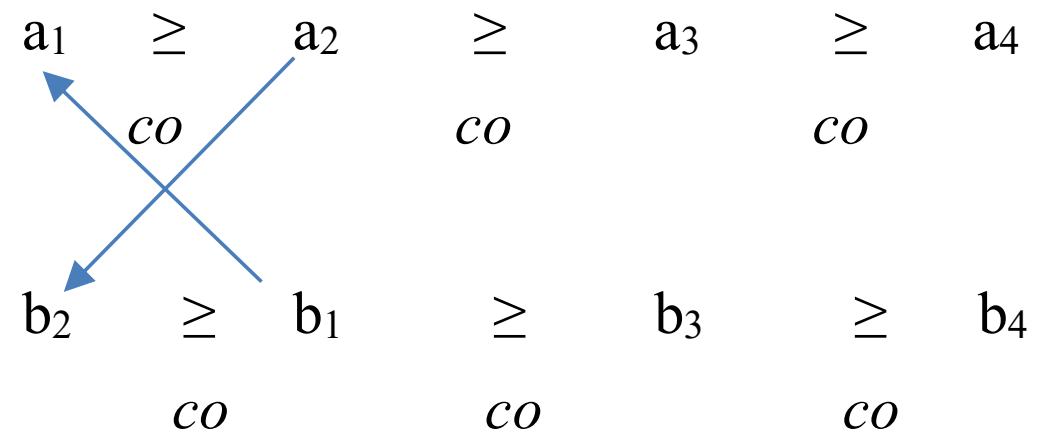

**Proof of (i):** The state $a_1$ is from CMPM $M_1$ and state $b_2$ belongs to CMPM $M_2$ and $n = 4$, the number of CFSMs and their corresponding unfoldings. The vector $(a_2, a_3, a_4)$ is the environment vector of $a_1$ and similarly $(b_1, b_3, b_4)$ is the environment vector of the state $b_2$. It is interesting to note that all the components from $a_1$ to $a_4$ are pairwise concurrent and are also related by causality, as shown above. Similarly the components $b_1$ to $b_4$.

Now, it is given that $(a_1 \, co \, b_2)$. By definition of concurrent paths, $a_1$ to $a_4$ and $b_2$ to $b_4$, no two states of these path segments are in conflict. Thus $a_1$ and $b_1$ can only be related by *seq* and so can $a_2$ and $b_2$. If $(b_2 \; seq \; a_2)$ were to be true, $(b_2 \; seq \; a_1)$ will be the consequence, contradicting $(b_2 \, co \, a_1)$. Thus $(a_2 \; seq \; b_2)$ will be the case, which is nothing but, $((a_2 = env_{12}) \; seq \; b_2)$. Similarly it can be proved that $(b_1 \; seq \; a_1)$, which is nothing but $((b_1 = env_{21}) \; seq \; a_1)$.

**Proof of (ii):** It is given that, $(a_2 \; seq \; b_2) \wedge (b_1 \; seq \; a_1)$. We need to show that $(a_1 \, co \, b_2)$. This is trivially true because, $(a_2 \; seq \; b_2)$ implies that, $a_2$ exits and entry of $b_2$ fills the place of $a_2$ which was in concurrence with $a_1$. Thus, $(a_1 \, co \, b_2)$ holds. Similarly we can show from $(b_1 \; seq \; a_1)$, $(b_2 \, co \, a_1)$ is true. Hence the result. In addition, $\neg(a_3 \; conf \; b_3)$ and $\neg(a_4 \; conf \; b_4)$ can be proved by contradiction since if they were in conflict, $(a_2 \; conf \; b_2)$ and $(a_1 \; conf \; b_1)$ will result by causal dependence, $(a_2 \geq a_3 \geq a_4)$ and $(b_2 \geq b_1 \geq b_3 \geq b_4)$ in addition to being pairwise concurrent. We exploit this property in model-checking.

**Property 2** Restricted transitivity of the *co* relation:

If $a_i \, co \, a_j$, and $a_j \, co \, a_k$, then $a_i \, co \, a_k$ where $a_i$, $a_j$, $a_k$ are respectively the states of unfoldings $M_i$, $M_j$ and $M_k$ respectively, $i \neq j \neq k$.

**Proof:**

Consider the following example with 4 components i.e, $n$=4

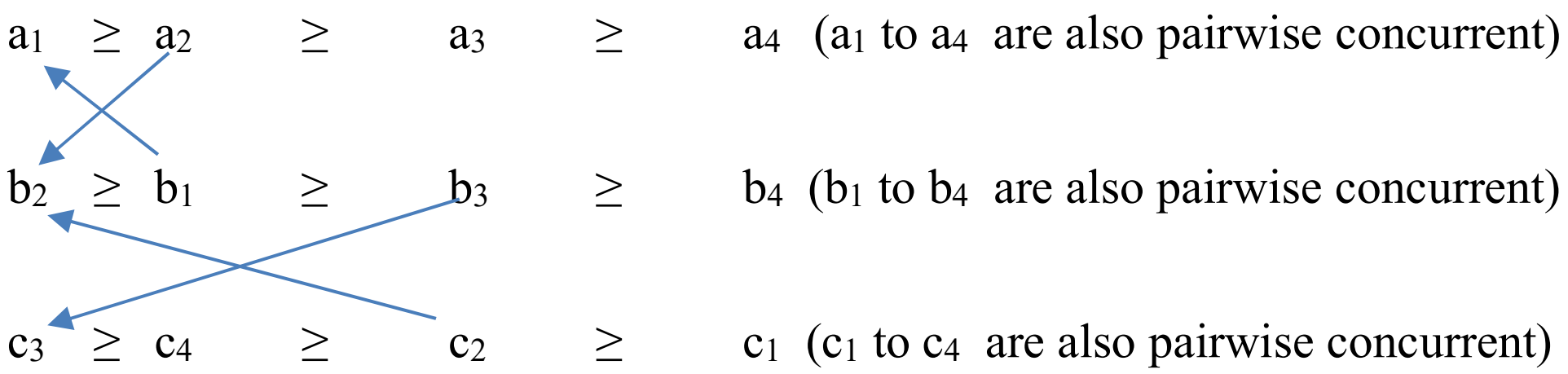

The arrows indicate sequential dependency, with $(a_2 \; seq \; b_2)$, $(b_1 \; seq \; a_1)$ , $(b_3 \; seq \; c_3)$ and $(c_2 \; seq \; b_2)$ since it is given that, $(a_1 \, co \, b_2)$ and $(b_2 \, co \, c_3)$. We need to show that $(a_1 \, co \, c_3)$.

In order to show that $(a_1 \, co \, c_3)$, using **property 1**, it is enough to show that $(a_3 \; seq \; c_3)$ and $(c_1 \; seq$

$a_1$). Now, it is true that $b_1$ and $c_1$ cannot be in conflict because if they were in conflict, it would propagate to make ($b_2$ *conf* $c_3$), which is not the case. Also ($b_1$ *seq* $c_1$) is not possible, since in that case, together with ($c_2$ *seq* $b_2$), and ($c_2 \geq c_1$) it would follow that ($b_1$ *seq* $b_2$) which is a contradiction because ($b_2$ *co* $b_1$). Therefore, ($c_1$ *seq* $b_1$) is the only possibility and since ($b_1$ *seq* $a_1$) is true, ($c_1$ *seq* $a_1$) is true. Very similarly it can be shown that, if ($b_3$ *seq* $a_3$) were to be true, it would violate ($b_3$ *co* $b_2$) and so it is only true that ($a_3$ *seq* $b_3$) and since ($b_3$ *seq* $c_3$), it follows that ($a_3$ *seq* $c_3$). Thus, it follows from ($a_3$ *seq* $c_3$) and ($c_1$ *seq* $a_1$) that ($a_1$ *co* $c_3$), by **property 1**.

## 4. Set of unfolded CFSMS is a Sum machine as opposed to the conventional Product machine of CFSMs

The set of CFSM unfoldings represent a concurrent set of Kripke structures. Together, they constitute a sum machine comprising the union of all the states of the set of unfolded trees. By virtue of its construction, the sum machine simulates the entire product machine by randomly and sequentially moving across one component unfolded CFSM to the other, through the synchronization points.

**Definition 6** A *sum machine* consisting of *n* CMPMs (unfolded CFSMs) can be defined as below:

($M_{i,i=\{1..n\}}$ , $\leq$, *seq, conf, co*) where,

- $M_i$ is i^th component of the set of CMPMs, $\forall i = \{1..n\}$,
- $\leq$ is the global causaliy order,
- *seq* is the global sequence relation,
- *conf* is the global conflict relation and,
- *co* is the concurrency relation.

### 4.1 Mapping between Product machine and Sum Machine

**Definition 7** A *configuration s* of sum machine is a vector of *n* CMPM states ($s_1, s_2, ...s_n$) such that the component states are pairwise concurrent. That is, ($s_i$ *co* $s_j$) for all $i,j \in \{1..n\}, i \neq j$.

The initial configuration of sum machine is $s_0 = (s_{01}, s_{02}..., s_{0n})$ , the vector of all the *n* initial states of unfolded CFSMs. These initial states are by default pairwise synchronous output states to begin with. That is, ($s_{0i}\ Rsync_{0i}\ s_{0j}$), for all i,j $\in \{1..n\}, i \neq j$. Also, $env_i(s_{0i}) = (s_{01}, s_{02}..., s_{0n})$, $i \in \{1..n\}$. Thus the initial configuration is the same as the env-vector of all the initial states that are pairwise synchronous and so trivially concurrent.

The set of *static configurations* of sum machine is the set of all *env-vectors* of all the component CMPMs, since they are statically stored. The set of all configurations that are not statically represented are *dynamic configurations*, reachable on need basis dynamically, depending on the property to be verified.

#### 4.1.1 Paths (of product machine) and concurrent set of local paths (of sum machine) correspond respectively to Interleavings and Runs

Consider a *concurrent local path set* of the sum machine $\Pi = (\Pi_1, \Pi_2, ... ,\Pi_n)$ with $\Pi_1 = (s_{01}\ R_1\ s^1{}_1\ R_1\ s^2{}_1, ....)$ , $\Pi_2 = (s_{02}\ R_2\ s^1{}_2\ R_2\ s^2{}_2, ....)$ ..., $\Pi_n = (s_{0n}\ R_n\ s^1{}_n\ R_n\ s^2{}_n, ....)$. Without loss of generality, let us assume all transitions of all the paths above are asynchronous (purely local). The initial, global state-vector is given by $s_0 = (s_{01}, s_{02}, ...s_{0n})$.

Let us now define a *global path P* of the state-vectors generated from concurrent local path set $\Pi$ as follows: P = ($s_0$ $R_1$ $s^1$ $R_2$ $s^2$ $R_3$ ... $R_n$ $s^n$…) where, $s^1 = (s^1{}_1, s_{02}, ...s_{0n})$ , $s^2 = (s^1{}_1, s^1{}_2, ...s_{0n})$, ...$s^n = (s^1{}_1, s^1{}_2, ...s^1{}_n)$., and $R = \bigcup_{i=\{1..n\}} R_i$. *P* is the global path traversed by the successive configurations with the initial configuration $s_0$. The first transition is that of CMPM $M_1$ of *R* denoted by $R_1$, the second is from $M_2$ represented by $R_2$ and so on until the $n^{th}$ transition from $M_n$ by $R_n$. The superscript denotes the position of the configuration from the initial one and the subscript denotes the index of the unfolded CFSM whose transition is made. The global path P represents the path traversed by the configurations starting from $s_0$ and making successive transitions of all *n* unfolded CFSMs in the order of their indices, $i = \{1..n\}$. In general, global paths can be formed by traversing the *n* unfolded CFSMs in any arbitrary order. Thus a given concurrent local path set $\Pi$ can be used to generate/trace multiple global paths depending on the order in which the component unfoldings are traversed. Each global path traced corresponds to an *interleaving* of the *run* corresponding to the concurrent path set $\Pi$. The size of conflict relation $|conf|$ decides the number of *runs* and the size of concurrency relation $|co|$ , controls the number of *interleavings* of a given *run.*

We do depth-first search of all the *n* component unfoldings of the sum machine in parallel to find instances of local reachable state components. The next step is to show that those states are concurrent to each other. Two states $s_i$ and $s_j$ belonging to CMPMs $M_i$, $M_j$ respectively can be checked for their *concurrency* by testing if their respective env-vector components *i* and *j* are reachable from each other, (in the sense that $s_i$ must be reachable from $env_{ji}(s_j)$ and $s_j$ from $env_{ij}(s_i)$).

### 4.2. Bisimulation of Product machine and the Sum machine

#### 4.2.1. Equivalence classes of configurations of Sum machine

Consider an unfolded CFSM state $s_i$ of $M_i$. The set of all configurations with $s_i$ as their $i^{th}$ component form an equivalence class whose representative is $env_i(s_i)$ which is a static configuration. Thus there are as many equivalence classes as there are sum machine states. [ $env_i(s_i)$] is the equivalence class of $env_i(s_i)$ given by all the reachable configurations *s*, whose $i^{th}$ component is $s_i$. Using these equivalence classes we can define *bisimulation equivalence* between product machine and sum machine and also between sum machine and its *quotient transition system.*

**Definition 8** The transition system of CMPM $M_i$ is defined as $TS(M_i) = (S_i, R_i, AP_i, L_i, s_{oi})$ where,

- $S_i$ is the set of local states of CMPM $M_i$,
- $R_i$ is the local transition relation,
- $AP_i$ is the set of atomic propositions,
- $L_i : S_i \rightarrow 2^{APi}$ is the labelling function mapping atomic propositions to each state, and
- $s_{oi}$ is the initial state.

Based on configurations and transitions among them, a single transition system corresponding to a product machine simulated by the sum machine can be defined as follows :

**Definition 9** The transition system of *sum machine M* is defined as $TS(M) = (S, R, AP, L, s_o)$ where,

- $S$ is the set of all configurations,
- $R = \bigcup_{i=\{1..n\}} R_i$ is the transition relation,
- $AP = \bigcup_{i=\{1..n\}} AP_i$ is the set of all atomic propositions,
- $L : S \rightarrow 2^{AP}$ is the labelling function mapping atomic propositions to each configuration, and
- $s_o$ is the initial configuration.

**Definition 10** The conventional transition system of the *product machine* of given set of CFSMs is given by $TS(M_f) = (S_f, R_f, AP_f, L_f, s_{0f})$ *where,*

- $S_f$ is the set of all CFSM state vectors
- $R_f \subseteq S_f \times S_f$ is the transition relation among state vectors,
- $AP_f = \times_{i=1..n} AP_{fi}$ is the set of all atomic propositions,
- $L_f : S_f \rightarrow 2^{APf}$ is the labelling function mapping atomic propositions to each configuration,

and

- $s_{of}$ is the initial CFSM-state vector

**Definition 11** Consider two transition systems TS1= ($S_1, R_1, AP_1, L_1, s_{o1}$) and TS2 = ($S_2, R_2, AP_2, L_2, s_{o2}$). A relation $\sim \subseteq S_1 \times S_2$ is a *bisimulation relation* iff: ($s_{o1} \sim s_{o2}$) and,

$\forall (s_1 \sim s_2)$, it holds: $L_1(s_1) = L_2(s_2)$ and,

if ($s_1$ $R_1$ $s'_1$) then ($s_2$ $R_2$ $s'_2$) where ($s'_1 \sim s'_2$) and,

if ($s_2$ $R_2$ $s'_2$) then ($s_1$ $R_1$ $s'_1$) where ($s'_1 \sim s'_2$).

Two transition systems *TS1* and *TS2* are *bisimilar* denoted $TS_1 \sim TS_2$ iff: there is a *bisimulation relation* between them.

It can be shown that the transition system of product machine, *TS(Mf)* and *TS(M)*, the transition system of sum machine are *bisimilar* by *construction* and induction since there is a *bisimulation* relation $\sim$ between them as follows:

$\sim \subseteq S \times S_f$ such that ($s \sim s_f$) and,

$\forall (s \sim s_f)$, it holds: $L(s) = L_f(s_f)$ and,

if ($s$ $R$ $s'$) then ($s_f$ $R_f$ $s'_f$) where ($s' \sim s'_f$) and,

if ($s_f$ $R_f$ $s'_f$) then ($s$ $R$ $s'$) where ($s'_f \sim s'_f$).

#### 4.2.2 The Bisimulation Quotient transition system of sum machine

The *quotient transition system* of sum machine TS(M /~) = (S/~, R/~, AP, L/~, $s_0/_\sim$) where, S/~ is the set of equivalence classes with arbitrary members [s] and [s'], R/~ is *defined* as: [s] R/~ [s'] iff: s R s' such that [s], [s'] ∈ S/~ , s ∈ [s], s'∈ [s']. L/~ is defined as L/~([s]) = L(s).
and $s_0/_\sim = [s_0]$.

The quotient transition system of sum machine TS(M/~) is what is generated *statically* using which the rest of the states and transitions of sum machine TS(M) can be dynamically generated on need basis depending on the requirements of reachability analysis.

## 5. Syntax and Semantics of CDTL logic over sum machine and CTL over product machine

Backus-Naur form of CDTL Logic Syntax is the following:

$\phi_i ::= p_i \mid \neg\phi i \mid \phi i \wedge \phi i \mid \phi i \vee \phi i \mid \phi i \rightarrow \phi i \mid A_iX i\phi i \mid E_iX i\phi i \mid A_iF i\phi i \mid E_iF i\phi i \mid A_iG i\phi i$
$\mid E_iG i\phi \mid A_i[\phi i \; U \; \phi i] \mid A_i[\phi i \; U \; \phi i]$.

*Semantics of Local CDTL formulas over sum machine components*

- $M_i, s_i \models p_i$ iff: $p_i \in L_i(s_i)$
- $M_i, s_i \models \neg\phi_i$ iff: $M_i, s_i \not\models \phi_i$
- $M_i, s_i \models \phi_i \wedge \psi_i$ iff: $(M_i, s_i \models \phi_i) \wedge (M_i, s_i \models \psi_i)$
- $M_i, s_i \models \phi_i \vee \psi_i$ iff: $(M_i, s_i \models \phi_i) \vee (M_i, s_i \models \psi_i)$
- $M_i, s_{i1} \models A_iF_i\phi_i$ iff: $\forall \pi_i = (s_{i1} \; R_i \; s_{i2} \; R_i \; s_{i3} \; R_i ... s_{ik} \; R_i \; s_{ik+1}, ...) \; M_i, s_{ik} \models \phi_i$
- $M_i, s_{i1} \models E_iF_i\phi_i$ iff: $\exists \pi_i = (s_{i1} \; R_i \; s_{i2} \; R_i \; s_{i3} \; R_i ... s_{ik} \; R_i \; s_{ik+1}, ...) \; M_i, s_{ik} \models \phi_i$

- $M_i, s_{i1} \models A_i[\phi_i \, U \, \psi_i]$ iff: $\forall \pi_i = (s_{i1} \, R_i \, s_{i2} \, R_i \, s_{i3} \, R_i \ldots s_{ik} \, R_i \, s_{ik+1}, \ldots) \; M_i, s_{ik} \models \psi_i$ and $\forall j < k$

  $M_i, s_{ij} \models \phi_i$

- $M_i, s_{i1} \models E_i[\phi_i \, U \, \psi_i]$ iff: $\exists \pi_i = (s_{i1} \, R_i \, s_{i2} \, R_i \, s_{i3} \, R_i \ldots s_{ik} \, R_i \, s_{ik+1}, \ldots) \; M_i, s_{ik} \models \psi_i$ and $\forall j < k$ $M_i, s_{ij} \models \phi_i$

Backus-Naur form of CTL Logic Syntax is the following:

$\phi ::= p_i \mid \neg\phi \mid \phi \wedge \phi \mid \phi \vee \phi \mid \phi \rightarrow \phi \mid AX\phi i \mid EX\phi \mid AF\phi \mid EF\phi \mid AG\phi$
$\mid EG\phi \mid A[\phi \, U \, \phi] \mid A[\phi \, U \, \phi]$.

*Semantics of Local CTL formulas over product machine*

- $M_f, s_f \models p_f$ iff: $p_f \in L_f(s_f)$
- $M_f, s_f \models \neg\phi$ iff: $M_f, s_f \not\models \phi_i$
- $M_f, s_f \models \phi \wedge \psi$ iff: $(M_f, s_f \models \phi) \wedge (M_f, s_f \models \psi)$
- $M_f, s_f \models \phi_i \vee \psi$ iff: $(M_f, s_f \models \phi) \vee (M_f, s_f \models \psi)$
- $M_f, s_{f1} \models AF\phi$ iff: $\forall \pi = (s_{f1} \, R_f \, s_{f2} \, R_f \, s_{f3} \, R_f \ldots s_{fk} \, R_f \, s_{fk+1}, \ldots) \; M_f, s_{fk} \models \phi$
- $M_f, s_{f1} \models EF\phi$ iff: $\exists \pi = (s_{f1} \, R_f \, s_{f2} \, R_f \, s_{f3} \, R_f \ldots s_{fk} \, R_f \, s_{fk+1}, \ldots) \; M_f, s_{fk} \models \phi$
- $M_f, s_{f1} \models A[\phi \, U \, \psi]$ iff: $\forall \pi = (s_{f1} \, R_f \, s_{f2} \, R_f \, s_{f3} \, R_f \ldots s_{fk} \, R_f \, s_{fk+1}, \ldots) \; M_f, s_{fk} \models \psi$ and $\forall j < k$

  $M_f, s_{fj} \models \phi$

- $M_f, s_{f1} \models E[\phi \, U \, \psi]$ iff: $\exists \pi = (s_{f1} \, R_f \, s_{f2} \, R_f \, s_{f3} \, R_f \ldots s_{fk} \, R_f \, s_{fk+1}, \ldots) \; M_f, s_{fk} \models \psi$ and $\forall j < k$

  $M_f, s_{fj} \models \phi$

### *5.1 Deduction of Global formulas of CTL's product machine from Local Formulas of CDTL's sum mchine*

Following equivalences between global and local formulae follow from the definition of *configurations*, global paths, *concurrent local path sets* and their relationship as explained in a previous subsection.

- $M, s \models q \leftrightarrow \bigwedge_{i=\{1..n\}} (M_i, s_i \models q_i)$ where, $(s_i \text{ co } s_j)$, $i \neq j$, $i,j=\{1..n\}$.
- $M, s \models AX(\bigwedge_{i=\{1..n\}} q_i) \leftrightarrow \bigwedge_{i=\{1..n\}} (M_i, s_i \models A_i X_i q_i)$, where $(s_{oi} \, R_i \, s_i)$, such that $Mi, s_i \models q_i$ $\forall i = \{1..n\}$ and $(s_1 \text{ co } s_2 \text{ co} \ldots s_n)$.
- $M, s \models AF(\bigwedge_{i=\{1..n\}} q_i) \leftrightarrow \bigwedge_{i=\{1..n\}} (M_i, s_i \models A_i F_i q_i)$, where $(s_{oi} \, R_i^{+} \, s_i)$, $s_i \models q_i$ $\forall i = \{1..n\}$ and $(s_1 \text{ co } s_2 \text{ co} \ldots s_n)$.

From the above equivalences, ($\leftrightarrow$ is the equivalence operator) it follows that in order to check the property of a global product machine formula, it is enough to search the corresponding component CMPM trees depending on the property to be verified, and then check the concurrency of local states to deduce the global property.

### 5.2 Global Reachability from Local Reachabilities with parallel and distributed Algorithm

We do depth-first search of all the *n* component unfoldings of the sum machine in parallel to find instances of local components of the state *s* whose reachability is to be analysed.
The next step is to show that those states are concurrent to each other. Two states $s_i$ and $s_j$ belonging to unfolded CFSMs $M_i$, $M_j$ respectively can be checked for their concurrency by testing if their respective environment-vector components are reachable from each other, anchored at $s_i$ and $s_j$ (in the sense that $s_i$ must be reachable from $env_{ji}(s_j)$ and $s_j$ from $env_{ij}(s_i)$).

***Step 1***. The procedure for *depth-first searches* of $M_i$, $\{i=1..n\}$ each, is same as the traditional product machine search[11] but can be done in parallel, with *n* processors, one for each unfolded CFSM (sum machine component). This step results in the complexity of *N*, the size of a component unfolding, which is given by $(d*N_f)$, where *d* is a constant, depending on the *interdependency/ degree of coupling* among the CFSMs given, dictated by the size of $|Rsync|$ relation and *conflict* due to non-determinism in the given CFSMs specification. We generally consider *loosely-coupled* CBS.

***Step 2.*** The next step is to find the concurrency among the locally reached states, found as a result of *step 1*. Checking of concurrency between two states $s_i$ and $s_j$, that is, checking if ($s_i$ *co* $s_j$) is satisfied, follows from **property 1**.
State $s_i$ must be reachable from $env_{ji}(s_j)$, the $i^{th}$ component of *environment vector* of $s_j$ and state $s_j$ must be reachable from $env_{ij}(s_i)$. If either of these conditions is not satisfied, it means that $s_i$ and $s_j$ must be in *sequence or conflict*, and *not concurrent*.

### 5.3 Discussion of Complexity of Distributed, Reachability Analysis

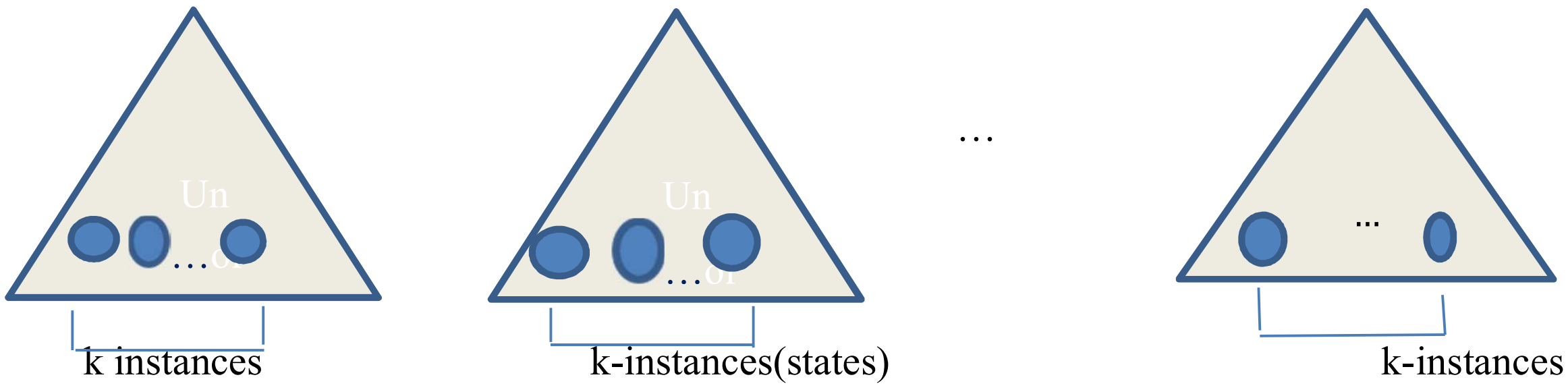

Unfolding $M_1$ Unfolding $M_2$ … Unfolding $M_n$

In each unfolding $M_i$, $i=\{1..n\}$ we find the local reachabilities of up to *k* states that are in conflict stemming from *k* different runs. Now for global reachability, we have to check the satisfaction of concurrency among these state instances that are mutually in conflict, each representing a run. We can map this problem to a satisfiability problem as follows say, with $n=4$ and $k=3$:

$(a_1 \lor b_1 \lor c_1) \land (a_2 \lor b_2 \lor c_2) \land (a_3 \lor b_3 \lor c_3) \land (a_4 \lor b_4 \lor c_4)$ where,

$a_1$, $b_1$, $c_1$ are the 3 states in *conflict*, representing 3 different *runs* satisfying the local reachability in $M_1$ and similarly $a_2,b_2,c_2$ in $M_2$ , $a_3,b_3,c_3$ in $M_3$ and $a_4,b_4,c_4$ in $M_4$.

Each comparison involves *2logN* operations where *logN* is the height of the unfolded tree, *N* being the number of states in the tree, *logN* representing the *maximum path length* to check the two reachabilities of the concerned states from their respective components of the environment vectors, to check their concurrency according to **property 1**. Total number of operations = $k^2 * 2logN=$

$O(k^2 logN)$ per tree.

### 5.3.1 Chain of Transitivity

The chain of transitivity follows from the concurrency between states $a_1$, $b_1$, $c_1$ of $M_1$ and states $a_2$, $b_2$ and $c_2$ of $M_2$ according to **property 2.** Similarly concurrency between $M_2$ and $M_3$. Put together, in general, the chain of transitivity applied $n$ times to give rise to the conjunction ( $x_1 \wedge x_2 \wedge x_3 \wedge \ldots \wedge x_n$) where $x_i$ represents one or more of $a_i$, $b_i$ and $c_i$, i={1..n}. Conjunction translates to concurrency just as conflicts to disjunction.

Total time complexity of ***Step 1*** and ***Step 2*** $= n*N + nk^2 logN$

$$= n*N + nk^2 logN$$
$$= O(n*N + nk^2 logN)$$

If we use $n$ processors in parallel, $O(N + k^2 logN)$ becomes the result.

### 5.3.2 Complexity of the Sum machine generation

If $N_f$ is the maximum number of states per CFSM and if there are $n$ interacting CFSMs the worst case size of the traditional synchronous product machine is $(N_f)^n$, which is *exponential* in the number of CFSMs. On the other hand, the size of sum machine/set of unfolded CFSMs in the worst case is only ($n*N_f*d$). Again using $n$ number of processors in *parallel* , we can generate the $n$ unfolded CFSMs comprising the sum machine in ($d*N_f$) time.

### 5.3.4. Expressiveness of CDTL

CDTL over sum machine adopts features of CTL over product machine by virtue of concurrent local path sets and global paths. In addition CDTL also adopts features of LTL over local path formulae so that nesting of linear-time operators $\mathbf{X}_i$, $\mathbf{F}_i$ and $\mathbf{U}_i$ are possible. The possibility of nested linear-time operators stems from the fact that we have the concept of sets of $n$-concurrent local paths of CMPMs. For instance, we can specify and verify the formula such as $\mathbf{A}((\mathbf{G}_1\mathbf{F}_1p_1 \wedge \mathbf{G}_2\mathbf{F}_2p_2 \ldots \wedge \mathbf{G}_n\mathbf{F}_np_n) \rightarrow \mathbf{F}q)$ where q is a fairness property and $p_1, p_2 \ldots p_n$ are local propositions of n CMPM/CFSM states using the concurrent set of global paths of configurations over $\Pi$. Here, identifying the satisfiability of $p_1, p_2 \ldots p_n$ can be done on paths $\Pi_1, \Pi_2, \ldots, \Pi_n$ respectively, given the sum machine structure. The formulas such as $\mathbf{AX}p$ corresponding to $(\mathbf{A}_1\mathbf{X}_1p_1 \wedge \mathbf{A}_2\mathbf{X}_2p_2 \wedge \ldots \wedge \mathbf{A}_n\mathbf{X}_np_n)$ can be specified and verified easily with the concurrent path-set $\Pi$. Nesting of branching-time operators such as in the formulae $\mathbf{AXEF}p$ , $\mathbf{E}_i\mathbf{F}_i\mathbf{A}_i\mathbf{G}_iq_i$ are possible as well, as in CTL*.

## 6. CONCLUSIONS AND FUTURE WORK

Given a CBS, specified as a set of $n$ CFSMs, as opposed to the traditional composition into a product machine which incurs a huge state-space explosion, we construct the sum machine, the quotient system of product machine whose synchronous state vectors alone are statically generated, and the rest are generable dynamically on need basis depending on the property to be verified.
We have proposed a *distributed version of branching-time logic* CDTL over sum machine structure using which model-checking can be performed efficiently. The model-checking complexity is ($n*N+nk^2 logN$) in the number of CFSMs $n$ with $N = d*N_f$, the maximum number of states per CFSM and $d$, a constant depending on degree of coupling, $|Rsync_{fi}|$ given in the specification. We can easily parallelize the model-checking algorithm with $n$ processors, thereby reducing the complexity to ($N + k^2 logN$). CTL model-checking can be encoded as a *satisfiability problem* which is proved to be NP-complete. But our polynomial complexity result makes us wonder if P = NP.

**Appendix**

**The Sum machine construction (Pseudo code)**

Algorithm:
Input: Set of *n* CFSMs specification in the form of *n graphs,* $F_i$, i={1..n}.
Output: Set of *n* unfolded CFSMs in the form of *n trees*, $M_i$, i = {1..n}

The algorithm given below unfolds the given CFSMs into unfolded trees by recursive simulation of CFSM states and transitions in parallel. Due to their inter-dependencies, every path generation of a given unfolded CFSM $M_i$ causally leads to the generation of certain paths of non-local CFSMs as well. Thus the unfoldings are generated as a group concurrently according to synchronization requirements instead of one at a time sequentially. The pseudocode mostly uses C style and Java style, for implementing the inter-thread ***waiting*** and ***notification***.

The main data structures are *env-vectors* and *waitlist*. The latter is a two-dimensional array with thread *i* adding into the *waitlist[i,j],* ∀*j=1..n*, its <state, transition> pairs, in sync with and read by thread *j*.

```
Main()
{
  Create and initialize n threads 1..n;
  for threads i= 1..n do in parallel
  {
      for j= {1..n , j≠i} do waitlist[i,j] := Null;
       /*waitlist of  thread i with all other threads
         initialized*/
      for all sfi ∈Sfi ins#(sfi)++ = 0;/*instance number
                              initialized*/
      si0 := <sfi0, 0)>;
      envi(si0) :=(s10, s20, ...sn0);
      Generate-unfolding(Fi, si);
   }/* for*/
}/*Main()*/

/*The pseudocode below simulates CFSM Fi  in global environment to generate unfolded CFSM
Mi in a depth-first recursive manner.*/

Generate_unfolding (Fi, si)
{
    if si is a cut-off state return; /* If si has an ancestor
          s'i such that fvec(envi(s'i)) = fvec(envi(si)) then si
          is cut-off state.*/
    for all transitions rfi=(fstatei(si), ai , s'fi) in Fi  do
    {
          s'i := Process-transition(si, rfi);
          Generate-unfolding(Fi, s'i);
     }
  }/*Generate-unfolding()*/

Process-transition(si, rfi)
{
    if  ai  of rfi is local/asynchronous action
            s'i := Gen-nextstate-async(si,  rfi);
```

```
    else if a_i is send/receive action in sync with
             thread j
        s'_i := Handle-nextstate-sync(s_i,  r_fi, j);
    return(s'_i);
}/*Process_transition()*/


Handle-nextstate-sync(s_i,  r_fi, j)
{
     if <s_j, r_fj> ∈ waitlist[j,i] /*If thread j has already
       added the partner <s_j, r_fj> in its waitlist[j, i]
       and is waiting for thread i for notification */
     {
         <s'_i , s'_j>:= Gen-nextstate-sync(s_i, s_j, r_fi, r_fj);
                Notify(thread j, s'_j); /* Thread j is notified by
                                  thread i to continue  from s'_j on */
     }
     else
     {
           waitlist[i,j] = waitlist[i,j] ∪ <s_i, r_fi>;
          /* Add <s_i, r_fi> in waitlist[i,j] */
          s'_i :=Wait(for thread j);  /* thread i  waits
                                                   for thread j for Notification*/
     }
     return(s'_i);
 }/*Handle-nextstate-sync()*/

 Gen-nextstate-async(s_i,  r_fi)
 {
          s'_i := <s'_fi, ins#(s_fi++)>;
          for k=1..n such that k ≠ i
                 env_ik(s'_i ) := env_ik(s_i);
          env_ii(s'_i) := s'_i;/* We generate s'_i and its env-vector*/
           R_i := R_i ∪ (s_i, s'_i);
               return(s'_i);
  }/*Gen-nextstate-async()*/

  Gen-nextstate-sync(s_i,s_j, r_fi,r_fj)
  {
       s'_i := <s'_fi, ins#(s'_fi++)>;
       s'_j :=<s'_fj, ins#(s'_fj++)>;
       for k=1..n such that k <> i,j
                             (env_ik(s'_i) =env_jk (s'_j)) := desc(env_ik(s_i) , env_jk(s_j)) ;
          /* The function desc returns the           descendent of the two states  in the local causality order
         R_k^+*/
       env_ii(s'_i) =env_ji (s'_j) :=s'_i;
       env_ij(s'_i) =env_jj (s'_j) :=s'_j;
       R_i := R_i ∪ (s_i, s'_i);
       R_j := R_j ∪  (s_j, s'_j);
       Rsync_i := Rsync_i ∪  (s'_i, s'_j);
                   Rsync_j := Rsync_j : = ∧  (s'_j, s'_i);
       return(<s'_i, s'_j>);
   }/* Gen-nextstate-sync()*/
```